\documentclass[nofootinbib,reprint, amsmath,amssymb,aps, superscriptaddress]{revtex4-2}

\usepackage{graphicx}
\usepackage{dcolumn}
\usepackage{bm}
\usepackage{aas_macros}
\usepackage{amsmath}
\usepackage{comment}
\usepackage[normalem]{ulem}
\usepackage{xcolor}

\begin{document}

\title{Pair luminosity and cooling of newborn strange star: Color-flavor-locked and two-flavor color superconducting quarks}

\author{Mikalai Prakapenia}
\affiliation{Institute of Physics, National Academy of Sciences of Belarus\\
220072 Nezalezhnasci Av. 68-2, Minsk, Belarus}

\author {Cheng-Jun Xia}
\affiliation{Center for Gravitation and Cosmology, College of Physical Science and Technology, Yangzhou University, Yangzhou 225009, China}

\author{Gregory Vereshchagin}
\affiliation{ICRANet, 65122 Piazza della Repubblica, 10, Pescara, Italy}
\affiliation{ICRA, Dipartimento di Fisica, Sapienza Universit\`a di Roma, Piazzale Aldo Moro 5, I-00185 Rome, Italy}
\affiliation{INAF -- Istituto di Astrofisica e Planetologia Spaziali, 00133 Via del Fosso del Cavaliere, 100, Rome, Italy}
\affiliation{ICRANet-Minsk, Institute of Physics, National Academy of Sciences of Belarus\\
220072 Nezalezhnasci Av. 68-2, Minsk, Belarus}
\date{\today}

\begin{abstract}
Following the previous work \cite{prakapenia2026} here we consider early thermal evolution of hot strange stars made of color superconducting quarks in two different pairing states: two-flavor color superconductor (2SC) and color-flavor-locked (CFL) phases, taking into account cooling by neutrinos and electron-positron pair creation due to the Schwinger process. We show that Schwinger luminosity in the electrosphere is a universal function of temperature, independent of quark matter phase for quark chemical potential $\mu_q>280$ MeV. The surface of a strange star in all the cases cools faster than its interior. This leads to a fast decrease of pair luminosity with time, so that it does not exceed $10^{46}$ erg/s at 1 second after strange star formation. Neutrino luminosity dominates over pair luminosity in all the cases except for the CFL phase with a large gap parameter $\Delta>60$ MeV, where there is a strong suppression of neutrino emission. The total energy emitted in electron-positron pairs is always smaller than the energy emitted in neutrinos, but they become comparable for the large gap parameter.

\end{abstract}

\maketitle

\section{Introduction}

One of the central questions in nuclear astrophysics is the state of matter in compact objects. The traditional picture is based on hadronic matter, but it is possible that at highest densities the strongly interacting matter forms a state containing deconfined up, down, and strange quarks \cite{1971PhRvD...4.1601B, 1984PhRvD..30..272W}, called strange quark matter (SQM). Compact astrophysical objects composed of SQM are called strange stars \cite{1986ApJ...310..261A, 2005PrPNP..54..193W}. Hybrid objects may exist as well, with a core of quark matter surrounded by hadronic matter \cite{1996csnp.book.....G}. Depending on the mass of strange quarks and the strength of pairing interactions, there may exist various color-superconducting phases for quark matter, such as the two-flavor color superconducting (2SC) phase and the color-flavor-locked (CFL) phase, which affects the properties and structures of strange stars~\cite{Alford2008_RMP80-1455,chen2026propertiesstrangequarkmatter}. In this paper, we consider cooling of strange stars with color superconducting 2SC and CFL quark phases. The 2SC phase is characterized by up and down quarks of two colors (red and green) participating in pairing, leaving blue color and strange quark unpaired \footnote{This phase is also termed "2SC+s" in the literature.}. In CFL phase quarks of all three colors and all three flavors are paired in antisymmetric pattern. 

A strange star can be formed as a result of phase transition in a neutron star \cite{2003ApJ...586.1250B,2004A&A...416..991A,2013PhRvD..87j3007P} or in a binary merger \cite{2022PhLB..83337388S}. In any case, a newborn strange star is very hot. The temperature in the stellar interior can be as high as $10^{11}$ K \cite{chengDai2001,1991ApJ...375..209H,2007ApJ...659.1519D}.  Powerful neutrino emission blows away the envelope of hadron matter, so the quark surface is bare and remains so as long as the temperature is higher than $\simeq10^7$ K \cite{Usov_1997, Usov_2001}. 
Neutron combustion was shown to produce a huge energy release of up to $10^{53}$ ergs. Thus, strange stars are considered as possible progenitors of explosive phenomena such as supernovae and gamma-ray bursts \cite{OLINTO198771, PhysRevD.111.063040,2003ApJ...586.1250B,Haensel:2006kkg} or fast radio bursts \cite{2021Innov...200152G}.

One of the unique features of strange stars is the existence on their surface of an electrosphere, namely a region with supercritical electric field exceeding the Schwinger limit for pair production $E_c=m_{e}^2c^3/\hbar e \simeq 1.3\times 10^{16}$ V/cm, where $m_{e}$ is the electron mass, $e$ is its charge, $c$ is the speed of light, and $\hbar$ is the reduced Planck constant.
An electrosphere of strange star exists because of the difference in strong and electromagnetic interactions on the surface. The density of the quark matter at the surface decreases on a length scale of a few fermis, set up by the strong interactions of quarks. Electrons are interacting with quarks via electromagnetic interactions and cannot be packed inside the quarks due to their quantum degeneracy pressure, so electron spatial distribution extends over the quark surface for hundreds of fermis.
In addition, the distribution of electrons is affected  by the sharp surface drop of quark density and depletion of $s$ quarks on the surface \cite{2005ApJ...620..915U}. This result holds both for unpaired quarks and for 2SC color superconducting phase. The electrosphere structure of CFL strange star has been considered in \cite{PhysRevD.70.067301} utilizing the results of \cite{madsen2000}, see, however \cite{OertelUrban2008}. In this paper, we consider color superconducting strange stars with an electrosphere. 

At low temperatures the electrosphere is static and does not produce electron-positron pairs due to the quantum degeneracy of electrons, i.e. the Pauli exclusion principle. In contrast, hot electrosphere produces a powerful pair wind \cite{1998PhRvL..80..230U,2024ApJ...963..149P} which may affect the cooling rate of the surface \cite{2021ApJ...922..214L,prakapenia2026}.

Observations of the cooling of compact objects is an important tool for identifying their nature \cite{2006NuPhA.777..497P}. There are several candidates of strange stars, notably the object HESS J1731-347 \cite{2026Univ...12...18N}.
Recall, that unlike quark stars, the surface of isolated neutron stars \cite{2001MNRAS.324..725G,2004ARA&A..42..169Y,2006NuPhA.777..497P} is cooling by photon emission. The cooling of quark stars is traditionally associated with neutrino emission \cite{2002PhRvL..89m1101P}. Recently \cite{prakapenia2026} we considered cooling of quark stars made of unpaired quarks taking into account both neutrino emission from the bulk and electron-positron emission from the surface and found that initially high luminosity in pairs cannot be sustained over long timescales of few seconds, due to insufficient thermal conductivity in the bulk. We have also shown there that neutrino trapping does not change the cooling behavior of the surface. Thus, focusing on thermal evolution of the surface one can neglect the effect of neutrinosphere and consider neutrino-transparent object even at very small timescales. 

In Section II we consider the structure of strange stars and use natural units with $\hbar = c = k_B = 1$, where $k_B$ is the Boltzmann constant, and the fine structure constant $\alpha = e^2 = 1/137$. In Section III we discuss thermal properties of strange stars and use the CGS system of units for clarity. In Section IV our numerical results are presented. The discussion and conclusions are presented in Section V.


\section{Surface and interior structure}
\label{hydroequil}
Various QCD-inspired models were proposed to describe SQM in the literature, e.g., perturbation model~\cite{Fraga2014_ApJ781-L25, Kurkela2014_ApJ789-127, Xu2015_PRD92-025025, Xia2017_NPB916-669, Xia2019_PRD99-103017}, linear sigma model~\cite{Holdom2018_PRL120-222001}, MIT bag model~\cite{Zhou2018_PRD97-083015, Miao2021_ApJ917-L22}, Dyson-Schwinger equations~\cite{Roberts1994_PPNP33-477, Alkofer2001_PR353-281}, equiparticle model~\cite{Peng2008_PRC77-065807, Xia2014_PRD89-105027}, quasiparticle model~\cite{Pisarski1989_NPA498-423, Schertler1997_JPG23-2051, Schertler1997_NPA616-659}, and Nambu-Jona-Lasinio model~\cite{Buball2005_PR407-205, Gholami2025_PRD111-103034}. In this work, we adopt  the MIT bag model, which determines the Grand potential per unit volume $\Omega$ of  quark matter~\cite{Fraga2001_PRD63_121702,Alford2005_ApJ629-969,Weissenborn2011_ApJ740-L14}. Below we consider electric and color neutral quark matter with beta-equilibrium condition. 

According to the bag model, the Grand potential per unit volume contains the bag energy per unit volume $B$. For the 2SC strange quark matter it is given by
\begin{gather}
\Omega_\text{2SC}
= -\frac{3a_4\mu_q^{4}}{4\pi^{2}} +\frac{3m_s^2\mu_q^2}{4\pi^{2}} \\ \notag
  - \frac{5 - 12\ln(m_s/2\mu_q)}{32\pi^{2}}m_{s}^{4}  \notag
  - \frac{\Delta^{2}\mu_q^{2}}{\pi^{2}}+B,
\end{gather}
where the QCD correction parameter $a_4$ is defined through the QCD coupling constant $\alpha_s$ as $ a_4 =1-2\alpha_s/\pi$~\cite{Fraga2001_PRD63_121702}. 

The first three terms of $\Omega_\text{2SC}$ are derived in Appendix A without QCD corrections when $a_4=1$, while the term with the gap parameter $\Delta$ represents the
contribution from color superconductivity. Another term $\Omega_e=-\mu_e^4/(12\pi^2)$ with electron chemical potential $\mu_e$ is small and can be neglected. We use the following relation between the electron chemical potential and the quark chemical potential $\mu_e=m_s^2/(2\mu_q)$ as described in Appendix A. 

The Grand potential per unit volume of CFL strange quark matter has the same structure as in the case of 2SC strange quark matter. It has the following form
\begin{gather}
\Omega_\text{CFL}
= -\frac{3a_4\mu_q^{4}}{4\pi^{2}} +\frac{3m_s^2\mu_q^2}{4\pi^{2}} \\ \notag
-\frac{1 - 12\ln(m_s/2\mu_q)}{32\pi^{2}}m_{s}^{4}
  - \frac{3\Delta^{2}\mu_q^{2}}{\pi^{2}}+B.
\end{gather}

The total pressure $P$ is defined as
\begin{gather}
P=-\Omega.
\end{gather}

The total pressure on the surface of the star equals zero. This condition determines the quark chemical potential with electron chemical potential for 2SC quark matter, while the electron chemical potential for CFL quark matter is zero.

To describe electron distribution on the surface we use Fermi-Dirac distribution function
\begin{equation}\label{FDdistr}
f_e= \frac{1}{1+\exp{\left[\left(\epsilon-\mu_e\right)/  T\right]}},
\end{equation}
with temperature $T$ and chemical potential $\mu_e$.

The electro-chemical equilibrium condition for electrons is
\begin{equation}\label{chemeq}
\mu_e= e \varphi,
\end{equation}
where $\varphi$ is the electrostatic potential. 

Introducing dimensionless electron degeneracy parameter $\eta\equiv \mu/T$ one can write the number density of electrons as
\begin{gather}\label{nedef}
n_e = - 2 \pi^{-2} T^3\text{Li}_{3}(-e^\eta), 
\end{gather}
where Li$_3$ is polylog function of the third order.

The Poisson equation for electrons near the surface is \cite{1986ApJ...310..261A,1995PhRvD..51.1440K}
\begin{equation}\label{poissoneq}
    \frac{d^2\varphi}{d z^2}=  4\pi \alpha  n_e(\varphi),
\end{equation}
where $z$ is the spatial coordinate normal to the surface. It is positive over the surface and negative under the surface.

Boundary conditions are the following:
\begin{gather}
\frac{d  \varphi }{d z}\bigg{|}_{ z=-0} =-
\frac{d \varphi }{dz}\bigg{|}_{ z=+0} = \frac{4\pi \alpha}{3} n_S, \\ \notag
\varphi \bigg{|}_{z = +\infty} = 0,~~~\varphi \bigg{|}_{z = -\infty} = \mu_e\bigg{|}_{P = 0}. 
\end{gather}
where $n_S$ is the surface density of $s$-quarks. The temperature $T$ is constant inside and outside the surface for both quarks and electrons.
Here we adopt the expression from \cite{XIA2017669}:
\begin{gather}
n_S = \int\limits_{0}^{p_{F_s}} \frac{6 p}{4\pi^2}\arctan\left(\frac{m_s}{p}\right)   dp 
\end{gather}
where $p_{F_s}= \sqrt{\mu^2_s - m_s^2}$. We apply this formula for both 2SC and CFL quarks neglecting the difference in quark chemical potentials ($\mu_s=\mu_q$) for 2SC and CFL strange quark matter.

\begin{figure}[ht]
\includegraphics[width=\columnwidth]{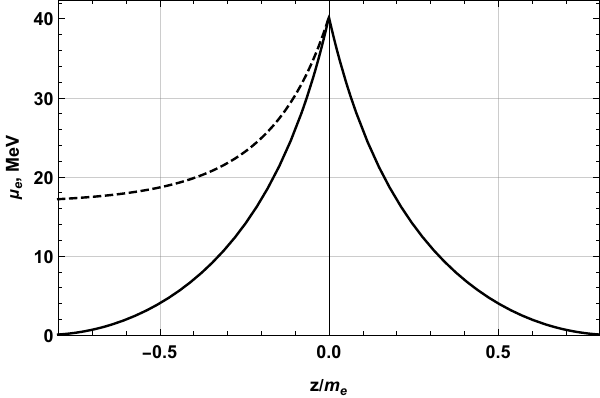}
\caption{Solution of Poisson equation for $T = 10$ MeV and $\mu_q =300$ MeV with $m_s=100$ MeV. Dashed: 2SC quarks. Solid: CFL quarks. The space coordinate is measured in units of the electron Compton wavelength.}
\label{PoissonSol}
\end{figure}

\begin{figure}[ht]
\includegraphics[width=\columnwidth]{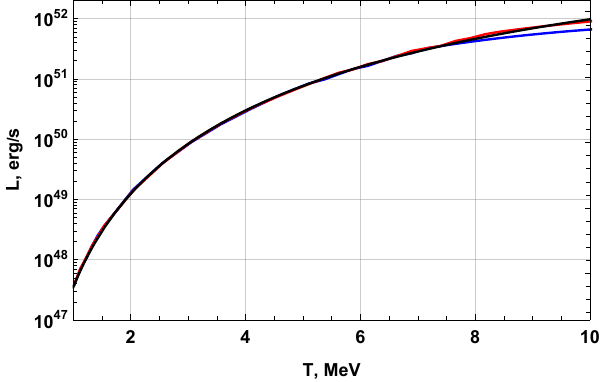}
\caption{Schwinger luminosity \eqref{Lschwinger} for $R=10$ km as a function of surface temperature $T$ for $\mu_q =350$ MeV (red) and $\mu_q =250$ MeV (blue) with $m_s=100$ MeV. Black curve represents fit formula \eqref{Lfit}.}
\label{Lfitplot}
\end{figure}

The solution of the Poisson equation near the surface of the star is shown in figure \ref{PoissonSol}. The depletion of $s$-quarks leads to the appearance of additional electronic states and the increase of the electron chemical potential in a thin surface layer of about one Compton wavelength of the electron. The electron density inside the CFL quark surface is zero, i.e. $\mu_e=0$, while for the 2SC quark surface $\mu_e=m_s^2/(2\mu_q)$.

To describe pair creation for non-vacuum initial state it is necessary to take into account Pauli blocking effect \cite{2023PhRvD.108a3002P,2023PhRvE.107c5204B}. We use the differential pair creation rate given in \cite{1987PhRvD..36..114G} with the additional Pauli blocking factor $(1-f_e)$. The rate is then given by an integral over the particle momentum $d^3p_e$ in the following form
\begin{gather}
\label{ndotschwinger}
\frac{dn_\pm}{dt} = - \frac{E/E_c}{2\pi^2} \int\limits_0^\infty d p~ p \left(1-f_e\right) \\ \notag \times\ln\left[  1-\exp{\left(-\frac{\pi(p^2+m_e^2)}{E/E_c} \right)}  \right]. 
\end{gather}

Finally, we estimate pair luminosity as
\begin{gather}\label{Lschwinger}
L_\text{Schwinger} = 4\pi R^2 \gamma m_e \int\limits_0^{z_0} \frac{dn_\pm}{dt} dz,
\end{gather}
where $\gamma$ is the Lorentz factor of positrons. 

To obtain luminosity in pairs we use the distribution function \eqref{FDdistr} in equation \eqref{ndotschwinger} where the chemical potential is given by the solution of the Poisson equation \eqref{poissoneq} for a given constant temperature.

The solution of the Poisson equation determines the Schwinger luminosity \eqref{Lschwinger}. We checked that for the temperatures $T$ ranging  from $0.5$ MeV to $10$ MeV Schwinger luminosity is unaffected by the change in parameter $m_s$ for the range of $m_s$ from $100$ MeV to $200$ MeV. For $\mu_q>280$ MeV Schwinger luminosity is independent on $\mu_q$ for $T<7.5$ MeV, and for $T>7.5$ MeV the luminosity increases up to a few percent. Thus, we can state that the energy flux of Schwinger pairs emitted from the surface of degenerate quark matter only depends on the surface temperature. We found that Schwinger luminosity \eqref{Lschwinger} can be fitted with the following formula, valid for $\mu_q>280$ MeV and $T<10$ MeV
\begin{gather}\label{Lfit}
L_\text{Schwinger} = 10^{48}\left(\frac{R}{10~\text{km}}\right)^2 \left(\frac{T}{1~\text{MeV}}\right)^{4.75}\\ \notag
\times\exp{\left(-\frac{T}{6~\text{MeV}}\right)}\exp{\left(-\frac{0.9~\text{MeV}}{T}\right)}~~\frac{\text{erg}}{\text{s}}. 
\end{gather}

Both eq. \eqref{Lfit} and Schwinger luminosity \eqref{Lschwinger} are shown in figure \ref{Lfitplot}.

\begin{figure}[t]
\includegraphics[width=\columnwidth]{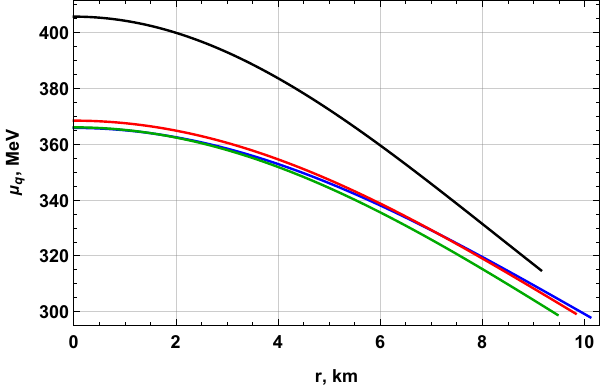}
\caption{Quark chemical potential as a function of radius of the star with $M=1.4M_\odot$. Blue: CFL quarks with $\Delta = 15$ MeV. Red: CFL quarks with $\Delta = 60$ MeV. Green: CFL quarks with $\Delta = 100$ MeV. Black: 2SC quarks with $\Delta = 60$ MeV.}
\label{internalmu}
\end{figure}

The quark density $n_{q}$ and the energy density $\epsilon$ can be found according to the basic thermodynamic relations:
\begin{gather}
n_q=-\frac{\partial\Omega}{\partial \mu_q},  \\  
\epsilon=\Omega+n_q \mu_q.
\end{gather}

In order for strange stars to be stable, the SQM hypothesis needs to be satisfied, i.e., SQM should be more stable than nuclear matter ($^{56}$Fe) with $3\epsilon/n_q <930$ MeV at $P=0$~\cite{1971PhRvD...4.1601B, 1984PhRvD..30..272W}, which restricts the choice of model parameters. In what follows, we fix $m_s=100$ MeV and $\alpha_s=0.1$. The gap parameter $\Delta$ ranges from $10$ MeV to $100$ MeV. We find the corresponding stellar configurations as a solution of the Tolman–Oppenheimer–Volkoff equation. We fix the chemical potential $\mu_q \simeq 300$ MeV at the surface and adjust the value of the bag constant $B$, so we have $B^{1/4}\simeq 150$ MeV. 

The quark chemical potentials inside the star with mass $M=1.4 M_\odot$ for both CFL strange quark matter and 2SC strange quark matter are shown in figure \ref{internalmu}. The radius of the star $R$ is about $10$ km for every chosen stellar configurations, and we conclude that the Schwinger luminosity $L_\text{Schwinger}$ is almost the same for every stellar configuration.

\section{Thermal conductivity, heat transfer and neutrino emissivity}
\label{neutrinotransport}

For the specific heat capacity $c_v$ of 2SC strange quark matter we assume that it is dominated by the ungapped quarks: blue up (bu), blue down (bd), red strange (rs), green strange (gs) and blue strange (bs).  We define the corresponding chemical potentials in Appendix A and use the standard expression for the heat capacity of relativistic degenerate fermions
\begin{gather}
c_v^\text{2SC} = \sum_{i=bu,bd,rs,gs,bs}\frac{1}{3}\frac{k_B^2 \mu_i^2}{\hbar^3c^3}T.
\end{gather}

The specific heat capacity  of the CFL strange quark matter was analyzed in \cite{Ferrer2021}. It was shown that the contribution of the massless Goldstone modes is the dominant one in the CFL phase. Therefore, we use the following expression for the heat capacity of superfluid phonons (which are massless relativistic bosons)
\begin{gather}
c_v^\text{CFL} = \frac{2\pi^2}{15}\frac{3^{3/2}k_B^4}{\hbar^3c^3}T^3.
\end{gather}

\begin{figure}[t]
\includegraphics[width=\columnwidth]{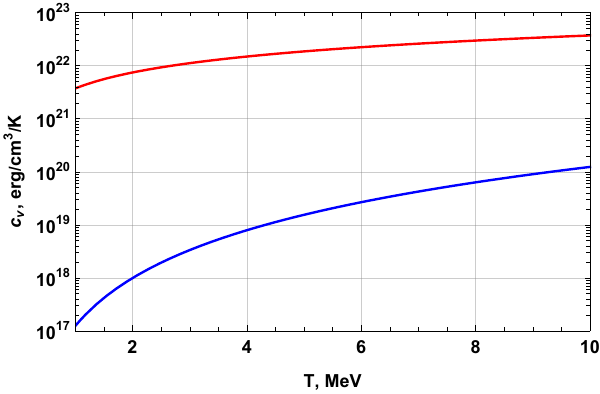}
\caption{Heat capacity of 2SC and CFL strange quark matter as a function of temperature for chemical potential $\mu_q=300$ MeV. Red: 2SC strange quark matter. Blue: CFL quark matter.}
\label{cVofT}
\end{figure}

The heat capacity of 2SC and CFL strange quark matter is shown in figure \ref{cVofT}. The heat capacity of CFL quark matter is more than two orders of magnitude smaller than that of 2SC quark matter at high temperatures $T\simeq10$ MeV. Neglecting neutrino emissivity and comparing only heat capacities one would find that CFL strange star cools down faster than 2SC strange star.

For 2SC strange quark matter we define a characteristic relaxation time for heat conductivity of a given particle type $\tau_i$. Some details of calculation of $\tau_i$ can be found in Appendix C. Following \cite{Shternin2022} the heat conductivity is
\begin{gather}
\kappa_\text{2SC} = \sum_{i=e,bu,bd,rs,gs,bs} \tau_i \frac{c^2 k_B^2 \pi^2 n_i T}{3\mu_i}.
\end{gather}

The heat conductivity of CFL quark matter was derived in \cite{braby2010}, and below we adopt the minimum value

\begin{gather}
\kappa_\text{CFL} =1.04\times10^{26}\left( \frac{\mu_q}{500\, \text{MeV}} \right)^8 \left( \frac{\Delta}{50\, \text{MeV}} \right)^{-6}~~\frac{\text{erg}}{\text{s cm K}}, 
\end{gather}

\begin{figure}[ht]
\includegraphics[width=\columnwidth]{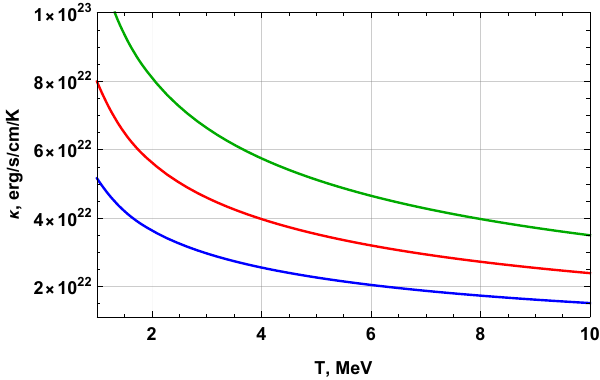}
\caption{Thermal conductivity of 2SC quarks as a function of temperature for selected values of chemical potential $\mu_q$. Blue: $\mu_q=300$ MeV. Red: $\mu_q=350$ MeV. Green: $\mu_q=400$ MeV.}
\label{k2SC}
\end{figure}

\begin{figure}[ht]
\includegraphics[width=\columnwidth]{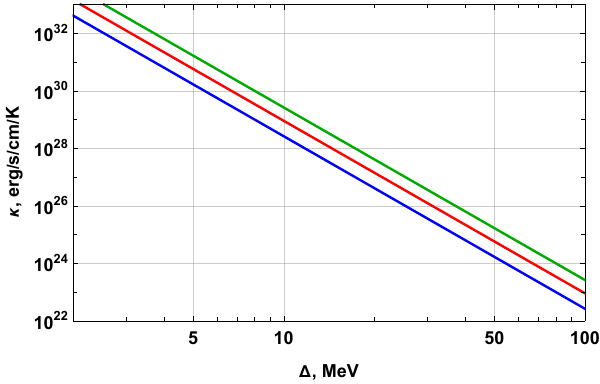}
\caption{Thermal conductivity of CFL quarks as a function of gap parameter for selected values of chemical potential $\mu_q$. Blue: $\mu_q=300$ MeV. Red: $\mu_q=350$ MeV. Green: $\mu_q=400$ MeV.}
\label{kappaCFL}
\end{figure}

We show the thermal conductivity in figures \ref{k2SC} and \ref{kappaCFL}. It is clear that the thermal conductivity of 2SC quarks decreases with increasing temperature, and it takes values of about $10^{22}$ erg/s/cm/K, which is similar to the case of unpaired quark matter. The thermal conductivity of CFL quark matter does not depend on temperature, but it strongly depends on the gap parameter $\Delta$. We consider the gap parameter in the range from $10$ to $100$ MeV, where the thermal conductivity of CFL quark matter ranges from $10^{22}$ erg/s/cm/K to $10^{30}$ erg/s/cm/K. Thus, for a small gap parameter, the thermal conductivity of CFL strange quark matter exceeds that of 2SC strange quark matter by several orders of magnitude. In \cite{prakapenia2026} we have shown that low thermal conductivity causes the instant formation of a steep temperature gradient at the stellar surface and a rapid decrease of the surface temperature.  So, based on considerations of thermal conductivity, one may expect that the surface of CFL strange star cools down slower than the surface of 2SC strange star. 

The temperature evolution is determined from the heat transfer equation with boundary conditions
\begin{gather}\label{heateq}
c_v \frac{\partial T}{\partial t} = - \frac{1}{4\pi r^2} \frac{\partial (4\pi r^2 F_r)}{\partial r} - Q_\nu,~~~F_r = -\kappa \frac{\partial T}{\partial r},   \\ 
\frac{\partial T(t,r=R)}{\partial r}= - \frac{F_R}{\kappa} =- \frac{L_\text{Schwinger}}{4\pi R^2 \kappa}, \label{boundary} \\  
T(t=0,r)=T_0, \label{initT}
\end{gather}
where $Q_\nu$ is the neutrino emissivity and $F_r$ is thermal flux density. The last equation expresses our assumption of an isothermal initial state of the star. Formation of a quark star is a complicated process, and the initial temperature distribution could be strongly nonuniform, as is the case for newborn neutron stars; see, e.g. \cite{2020ApJ...888...97B}. However, we adopt the initial condition \eqref{initT} for simplicity. 

The boundary condition on the surface is given by equation \eqref{boundary}, which means that the only energy flux from the surface is solely generated by the Schwinger process.

For 2SC strange quark matter we use the same expression as in unpaired strange quark matter, where neutrino emissivity is dominated by the direct URCA processes of unpaired blue $u$ and $d$ quarks. The energy emission per unit time and volume is  \cite{IWAMOTO19821}  
\begin{gather}
    Q^\text{2SC}_\nu = 3\times10^{24} \frac{\mu_{bu}\mu_{bd}}{(400\,\text{MeV})^2}\frac{\mu_e}{10\,\text{MeV}}T^6_9~~\frac{\text{erg}}{\text{s cm}^3}.
    \label{Qnu}
\end{gather}

We assume that neutrino emission due to direct URCA processes $d\rightarrow u + e + \bar\nu_e$ and $u+e\rightarrow d+\nu_e$ is absent in CFL quark matter, because of negligible electron fraction and strong quark pairing.  Since we consider cooling process on very short timescales (or, alternatively, at very high temperature) neutrino emission from Goldstone modes can be neglected. Instead, the dominant neutrino emission process is quark bremsstrahlung $q+q\rightarrow q+q+\nu+\bar\nu$, which is also exponentially suppressed due to quark pairing
\cite{IWAMOTO19821,JAIKUMAR2001345,Jaikumar2002}
\begin{gather}\label{Qnucfl}
Q^\text{CFL}_\nu = 10^{20}\left(\frac{n_q}{0.22~ \text{fm}^{-3}}\right)^{1/3}
T_9^8e^{-2\Delta/k_BT}~~\frac{\text{erg}}{\text{s cm}^3},
\end{gather}

\begin{figure}[ht]
\includegraphics[width=\columnwidth]{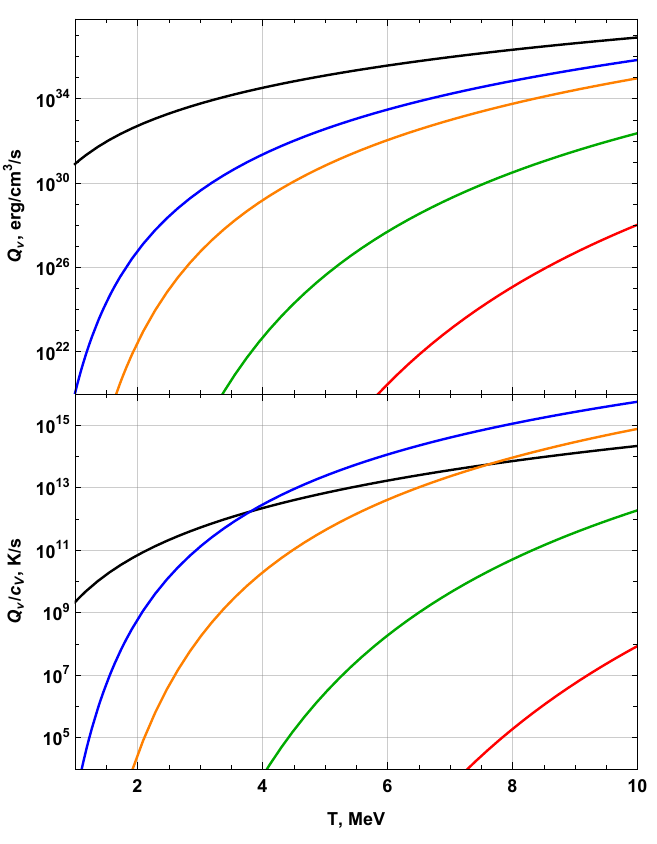}
\caption{Neutrino loss of CFL and 2SC strange quark matter as a function of temperature. Blue: CFL strange quark matter with $\Delta=10$ MeV. Orange: CFL strange quark matter with $\Delta=20$ MeV. Red: CFL quark matter with $\Delta=60$ MeV. Green: CFL strange quark matter with $\Delta=100$ MeV. Black: 2SC strange quark matter with $\Delta=60$ MeV and $\mu_q=350$ MeV.  }
\label{QnuofT}
\end{figure}

We show neutrino emissivity for 2SC and CFL strange quark matter in figure \ref{QnuofT}. As expected, neutrino emissivity of CFL quark matter is much smaller than that of 2SC quark matter. However, the ratio of neutrino emissivity and heat capacity $Q_\nu/c_v$, which determines the temperature evolution of the star, can be larger for CFL quark matter for gap parameters $\Delta \lesssim 20$ MeV. 


\section{Numerical results}

After considering basic thermal properties of strange quark matter in CFL and 2SC phases, we turn to the cooling process. We solved numerically the heat transfer equation for different stellar configurations presented in figure \ref{internalmu} that have a fixed mass $M = 1.4M_\odot$.
\begin{figure}[t]
\includegraphics[width=\columnwidth]{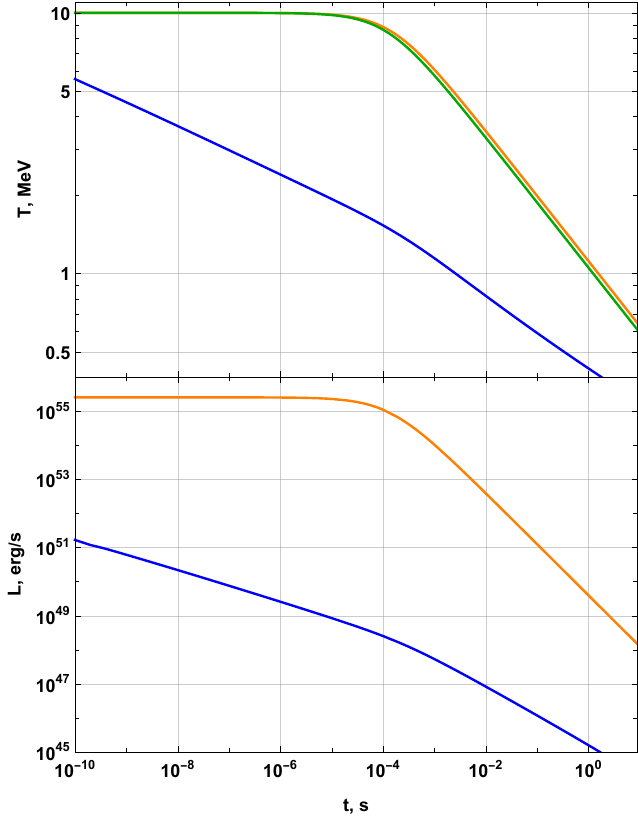}
\caption{Top: time evolution of the temperature for 2SC strange star with gap parameter $\Delta=50$ MeV at selected radii: surface (blue), center (orange) and one meter under the surface (green). Bottom: time
evolution of pair luminosity (blue) and neutrino luminosity (orange).}
\label{TLnuoft}
\end{figure}

First, we discuss the thermal evolution of the newborn 2SC strange star. The temperature evolution of the star is shown in figure \ref{TLnuoft} together with the luminosity evolution. The cooling proceeds in two stages. During the first stage, which lasts up to $\sim 10^{-4}$ s, internal temperature and neutrino luminosity remain constant. Thermal evolution at this stage, which is known in the cooling theory of neutron stars \cite{2020ApJ...888...97B}, strongly depends on initial temperature distribution in the star. However, given that neutrino luminosity strongly depends on the temperature, see equations (\ref{Qnu}) and (\ref{Qnucfl}), temperature gradients quickly smear out and the star forgets about its initial state. The second stage starts after $10^{-4}$ s and subsequent cooling dynamics becomes independent on the initial state, which justifies our initial isothermal approximation.

Both surface temperature and pair luminosity $L_\text{Schwinger}\propto t^{-1/2}$ decrease as power-laws in the first stage. After interior temperature of the star starts to decrease due to neutrino emission, the neutrino luminosity follows a different power law $L_\nu\propto t^{-3/2}$. At the same time, surface temperature evolution changes as well, as the amount of heat in the interior of the star decreases. This leads to gradual smearing of the temperature gradient at the surface (temperature values in the interior and at the surface become closer to each other with time) and to a faster decrease of pair luminosity $L_\text{Schwinger}\propto t^{-2/3}$. Neutrino luminosity greatly exceeds pair luminosity at all times. The total energy emitted in neutrinos at the time moment $t=1$ s is about $6.8\times10^{51}$ ergs, while the total energy emitted in pairs is only $8.5\times 10^{45}$ ergs. Later (when the temperature decreases below 0.5 MeV) the pair luminosity decreases exponentially \cite{2002PhRvL..89m1101P,prakapenia2026}.

It turns out that the cooling is similar to the case of a strange star made of unpaired quarks analyzed in \cite{prakapenia2026}. This is expected, as all basic thermal properties of 2SC strange quark matter are similar to those of unpaired strange quark matter. As we pointed out in the previous section, thermal conductivity of 2SC strange quark matter at the temperature $T=10$ MeV takes the value of about $10^{22}$ erg/s/cm/K, which is similar to the case of unpaired strange quark matter. But the temperature dependence of thermal conductivity is different, $\kappa_\text{unpaired}\propto T^{-1}$, see \cite{prakapenia2026}, while here we have $\kappa_\text{2SC}\propto T^{-1/2}$. In addition, the surface properties of quark matter, which lead to modification of the Schwinger luminosity at high temperatures, are taken into account in this work, but were neglected in \cite{prakapenia2026}. Both these changes affect the time evolution of the surface temperature, but the time evolution of the interior temperature is almost the same for both 2SC and unpaired quarks. 

\begin{figure}[t]
\includegraphics[width=\columnwidth]{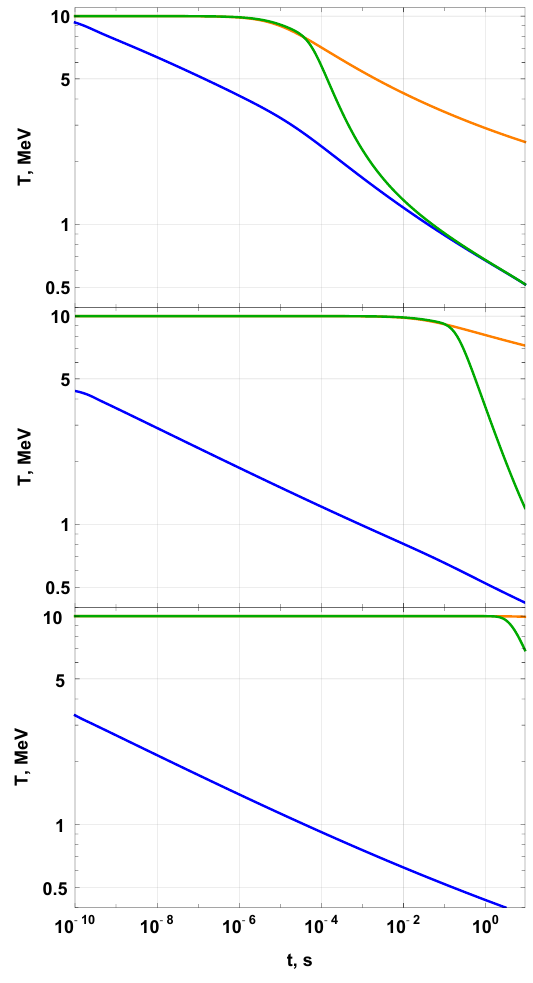}
\caption{Time evolution of the temperature of CFL strange star for selected gap parameter $\Delta=15$ MeV (top), $\Delta=60$ MeV (middle), $\Delta=100$ MeV (bottom) at different radii. Blue: surface temperature. Orange: central temperature.
Green: temperature at one meter under the surface.}
\label{Toft}
\end{figure}

\begin{figure}[t]
\includegraphics[width=\columnwidth]{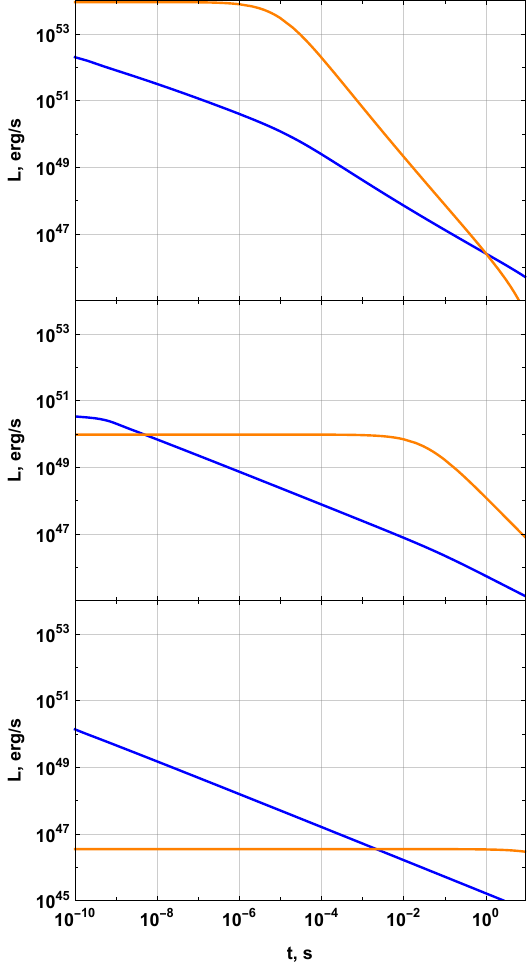}
\caption{Time evolution of pair luminosity
(blue) and neutrino luminosity (orange) of CFL strange star for selected gap parameter $\Delta=15$ MeV (top), $\Delta=60$ MeV (middle), $\Delta=100$ MeV (bottom).}
\label{Lnuoft}
\end{figure}

\begin{figure}[t]
\includegraphics[width=\columnwidth]{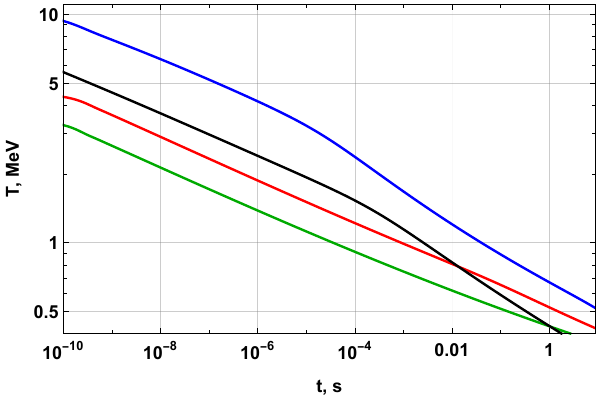}
\caption{Surface temperature evolution from figures \ref{Toft} and \ref{TLnuoft}. Blue: CFL strange star with $\Delta=15$ MeV. Red: CFL strange star with $\Delta=60$ MeV. Green: CFL strange star with $\Delta=100$ MeV. Black: 2SC strange star with $\Delta=60$ MeV.}
\label{Tsurfoft}
\end{figure}

Next, we discuss the thermal evolution of a newborn CFL strange star. The dependence of temperature on time is shown in figure \ref{Toft} and its luminosity evolution is shown in figure \ref{Lnuoft}.  We present three cases with increasing gap parameter $\Delta$:  $\Delta=15$ MeV (top), $\Delta=60$ MeV (middle) and $\Delta=100$ MeV (bottom). Both thermal conductivity and neutrino luminosity are high for a small gap parameter and decrease with increasing $\Delta$, as clearly seen in figure \ref{Toft}. Indeed, the surface temperature at the time moment $t=10^{-10}$ s becomes smaller when the gap parameter increases, as a consequence of reduced thermal conductivity. The duration of the first cooling phase, when interior temperature remains unchanged, strongly depends on the gap parameter. As discussed above, one can expect that the temperature gradients inside the star disappear at the end of this stage. In contrast,
the temperature gradient established at the surface layer is small for $\Delta=15$ MeV, and it increases for larger value of the gap parameter, due to decreasing thermal conductivity.

Both of these features can be clearly seen from the luminosity dependence on time, shown in figure \ref{Lnuoft}. Neutrino luminosity starts  decreasing at $t\simeq\{10^{-6},10^{-2},10\}$ s for $\Delta= \{15,60,100\}$ MeV. Pair luminosity also gets smaller with increasing gap parameter. At large gap parameter pair luminosity exceeds the neutrino luminosity, see the bottom figure \ref{Lnuoft}. Nevertheless, the total energy emitted by the time moment $t=1$ s is: $1.1\times10^{49}$ erg in neutrinos and $9.3\times10^{46}$ erg in pairs for $\Delta=15$ MeV;  $7.2\times10^{48}$ erg in neutrinos and $1.3\times10^{46}$ erg in pairs for $\Delta=60$ MeV; $3.7\times10^{46}$ erg in neutrinos and $3.0\times10^{45}$ erg in pairs for $\Delta = 100$ MeV.

Finally, we summarize these results in figure \ref{Tsurfoft}, where surface temperature is shown as a function of time for both 2SC and CFL strange stars. We note that the temperature of 2SC strange star is higher at early times than that of CFL strange stars with $\Delta=60$ MeV and $\Delta=100$ MeV, despite the fact that the 2SC strange star has much smaller thermal conductivity. Clearly, the cause is the difference in the specific heat capacities of the 2SC and CFL phases.

\section{Discussion and conclusions}

We considered early thermal evolution of a newborn strange star with color superconductivity taking into account both neutrino emission from the entire stellar body and pair creation in its electrosphere.  The cooling of the 2SC and CFL strange stars is reported, assuming an initial temperature of $10$ MeV.

By taking into account $s$ quark depletion at the surface we found that pair luminosity is independent of the phase of SQM if one neglects the difference in quark chemical potential or the charge screening of quarks.
Thus, Schwinger luminosity on the surface of a strange star is fully determined by the effect of $s$ quark depletion, which is almost independent of the phase of SQM.

It is clear from the results shown above that the cooling process of a newborn strange star crucially depends on the physical properties of three regions: the surface, the surface layer, and the interior of the star. The latter is cooling in our model by neutrino emission only. 

The surface thermal evolution is determined by the Schwinger process. Two additional factors play a role: heat transport from the interior (thermal conductivity) and the amount of heat remaining in a strange star (heat capacity and neutrino emission rate). These properties depend on the quark matter phase state, so in different states (unpaired quarks, 2SC or CFL) strange star thermal evolution is different. The interplay between strong cooling of the surface and insufficient thermal conductivity leads to the establishment of strong temperature gradients in a layer very close to the strange star surface, see \cite{prakapenia2026}. Here we confirm that this temperature gradient is present in all quark matter phases, and it is this effect which determines the rate of pair creation and consequently the luminosity in pairs.

The most interesting appears to be the dependence of pair and the neutrino luminosity on the gap parameter $\Delta$ of CFL quark matter. For larger $\Delta$ neutrino luminosity is suppressed, see eq. \eqref{Qnucfl}, resulting in prominence of the Schwinger process, so that the total energy emitted in pairs becomes comparable to the energy emitted in neutrinos.

In conclusion, we found that strong temperature gradients at the surface severely limit Schwinger luminosity in all phase states of quark matter. Electron-positron luminosity depends only on temperature, while neutrino luminosity also depends on the state of quark matter. In CFL phase with large gap parameter $\Delta$, the pair luminosity exceeds the neutrino luminosity and the total energy emitted in pairs becomes comparable to that of neutrinos.

\appendix
\section{Chemical potentials and charges of 2SC strange quark matter}

In Appendices A, B and C we derive thermal conductivity of 2SC strange quark matter following \cite{PhysRevC.90.055205,Alford_2010,Shternin2022}.

We order quark flavors as $(u,d,s)$ and quark colors as $(r,g,b)$ and write electromagnetic and color generators as follows
\begin{gather}
Q = \text{diag} \left( \frac{2}{3},-\frac{1}{3}, -\frac{1}{3}  \right), \\
T_3 = \text{diag} \left( \frac{1}{2},-\frac{1}{2}, 0  \right), \\
T_8 = \text{diag} \left( \frac{1}{3}, \frac{1}{3}, -\frac{2}{3}  \right),
\end{gather}
 which are  associated with the electric charge in $(u,d,s)$ space, color charges $Q_3$ and $Q_8$ in $(r,g,b)$ space. Note that there are totally eight color charges, but it is always possible to transform  to a gauge where all are zero except for $Q_3$ and $Q_8$. This is associated with the diagonal generators $T_3$ and $T_8$~\cite{Alford2008_RMP80-1455}.

The chemical potentials of quarks of different types can be expressed through  chemical potentials $\{\mu_q,~\mu_e,~\mu_3,~\mu_8\}$ utilizing the expression $\mu_i=\mu_q-Q\mu_e+T_3\mu_3+T_8\mu_8$ as
\begin{gather}
\mu_{ru} = \mu_q -\frac{2}{3}\mu_e + \frac{1}{3}\mu_3 + \frac{1}{3}\mu_8, \\
\mu_{gu} = \mu_q -\frac{2}{3}\mu_e - \frac{1}{3}\mu_3 + \frac{1}{3}\mu_8, \\
\mu_{bu} = \mu_q -\frac{2}{3}\mu_e - \frac{2}{3}\mu_8,\\
\mu_{rs} =\mu_{rd} = \mu_q +\frac{1}{3}\mu_e + \frac{1}{3}\mu_3 + \frac{1}{3}\mu_8, \\
\mu_{gs} =\mu_{gd} = \mu_q +\frac{1}{3}\mu_e - \frac{1}{3}\mu_3 + \frac{1}{3}\mu_8, \\
\mu_{bs} =\mu_{bd} = \mu_q +\frac{1}{3}\mu_e - \frac{2}{3}\mu_8, 
\end{gather}
where we have introduced the electron chemical potential $\mu_e$ for the electric charge, the color chemical potentials $\mu_3$ and $\mu_8$ for the color charges  $Q_3$ and $Q_8$.
We define the  Grand potential per unit volume of the 2SC phase as 
\begin{gather}
\Omega_{\text{2SC}} =
\frac{1}{\pi^{2}} \sum_{i=ru,gu,rd,gd} \int_{0}^{p_F^C} (p-\mu_{i})\,p^{2}dp
+ \\ \notag
\frac{1}{\pi^{2}} \sum_{i=bu,bd,bs,rs,gs} \int_{0}^{\sqrt{\mu_{i}^{2}-m_{i}^{2}}}
\left(\sqrt{p^{2} + m_{i}^{2}}-\mu_{i}\right) p^{2}\,dp \\ \notag
- \frac{1}{\pi^{2}}\Delta^{2}\mu_q^2,
\end{gather}
where $p_F^C = \mu_q - \frac{1}{6}\mu_{e} + \frac{1}{3}\mu_{8}$ is the average chemical potential of the Cooper pair. 

The electric and color neutrality conditions are
\begin{gather}
\frac{\partial \Omega_{\text{2SC}}}{\partial \mu_{e}}
= 0, \\
\frac{\partial \Omega_{\text{2SC}}}{\partial \mu_{3}}
= 0, \\
\frac{\partial \Omega_{\text{2SC}}}{\partial \mu_{8}}
= 0.
\end{gather}
An approximate solution for these conditions can be written as $\mu_3=0,~\mu_8=0,~\mu_e=m_s^2/(2\mu_q)$. Expanding $\Omega_\text{2SC}$ in powers of $m_s/\mu_q$ we obtain the result given in section II.

Then we introduce the broken/unbroken bases $X$ and $\tilde Q$ which are a linear combination of $Q$ and $T_8$:
\begin{gather}
\tilde Q = Q + \eta_1 T_8,~~~ X = T_8 - \eta_2 Q, \\ 
\eta_1=-\frac{1}{2},\\
\eta_2 = -\frac{e^2}{2g^2}=-2\tan^2\varphi = - \frac{\alpha}{2\alpha_s},
\end{gather}
 where $\varphi$ represents the mixing angle analogous to the Weinberg angle in the electroweak theory, $\alpha = e^2/(4\pi)$, $\alpha_s = g^2/(4\pi)$, $g$ is QCD gauge coupling parameter.

Finally, we define the new gauge couplings $e^{\tilde Q}$ and $e^X$:
\begin{gather}
e^{\tilde Q}= e \cos\varphi= \frac{2eg}{\sqrt{e^2+4g^2}} = \frac{4\pi \sqrt{\alpha\alpha_s}}{\sqrt{\pi\alpha+4\pi\alpha_s}}, \\
e^X= g \cos\varphi= \frac{2g^2}{\sqrt{e^2+4g^2}}=\frac{4\pi \alpha_s}{\sqrt{\pi\alpha+4\pi\alpha_s}}.
\end{gather}

The products of the coupling constant and the charge matrix for the ungapped fermions $(bu,bd,e,rs,gs,bs)$ are
\begin{gather}
Q_i^{T_8} = g \times  \left( -\frac{2}{3},-\frac{2}{3}, 0,\frac{1}{3},\frac{1}{3},-\frac{2}{3}  \right), \\
Q_i^Q = e \times \left( +\frac{2}{3},-\frac{1}{3}, -1,-\frac{1}{3},-\frac{1}{3},-\frac{1}{3}  \right), \\
Q_i^{\tilde Q} = e^{\tilde Q} \times  \left( 1,0, -1,-\frac{1}{2},-\frac{1}{2},0  \right), \\
Q_i^{X} = e^X \times  \biggl( \frac{2}{3}(-1+\frac{\alpha}{2\alpha_s}),-\frac{2}{3}(1+\frac{\alpha}{4\alpha_s}), \frac{\alpha}{2\alpha_s}, \\ \notag
\frac{1}{3}(1-\frac{\alpha}{2\alpha_s}),\frac{1}{3}(1-\frac{\alpha}{2\alpha_s}), -\frac{2}{3}(1+\frac{\alpha}{4\alpha_s}) \biggr). 
\end{gather}
We also write the average $Q^a_C$ for two quarks in a Cooper pair 
\begin{gather}
Q_C^{T_8} = \frac{g}{3}=\frac{\sqrt{4\pi \alpha_s}}{3},~~Q_C^Q = \frac{e}{6}= \frac{\sqrt{4\pi\alpha}}{6}, \\ \notag
Q_C^{\tilde Q} = 0,~~Q_C^X = \frac{g}{3 \cos\varphi}=\frac{\sqrt{\pi\alpha+4\pi\alpha_s}}{3}.
\end{gather}

\section{Matrix elements}
Consider the binary process $i+j\rightarrow i'+j'$. 
The energy and momentum transfers are $\omega$ and $\mathbf q$. We define $q \equiv |\mathbf q|$ and similarly for other momenta and velocities. The scattering matrix element for two incoming particles $i$ and $j$ with energies $\epsilon_i$ and $\epsilon_j$, momenta $\mathbf p_i$ and $\mathbf p_j$, velocities $\mathbf v_i$ and $\mathbf v_j$ has the following form
\begin{gather}
    |M_{ij}|^{2}=L_{l} \left| \sum_{a=\{T_8,Q\} }\frac{Q_i^a Q_j^a}{q^2+\Pi_l^{aa}} \right|^{2} \\ \notag
    + v_i^2 v_j^2 L_t \left| \sum_{a= \{X,\tilde Q\} } \frac{Q_i^a Q_j^a}{q^2-\omega^2+\Pi_t^{aa}} \right|^{2} \\ \notag
    - 2 v_i v_j L_{lt}\times \\ \notag
    \text{Re} \left[ \left( \sum_{a=\{T_8,Q\} } \frac{Q_i^a Q_j^a}{q^{2} + \Pi_l^{aa}} \right) \left( \sum_{a=\{X,\tilde Q\} } \frac{Q_i^a Q_j^a}{q^2-\omega^2+\Pi_t^{aa}} \right)^{*} \right],
\end{gather}
where
\begin{gather}
    L_{l} = \left(1 - \frac{q^{2} v_i^2}{4 p_i^2}\right)\left(1 - \frac{q^{2}v_j^2}{4 p_j^2}\right), \\ \notag
    L_{lt} = \left(1 - \frac{q^2}{4 p_i^2}\right)^{1/2}\left(1 - \frac{q^2}{4 p_j^2}\right)^{1/2}\cos\theta, \\ \notag
    L_{t} = \left(1 - \frac{q^{2}}{4 p_i^2}\right)\left(1 - \frac{q^{2}}{4 p_j^2}\right)\cos^{2}\theta + \frac{q^2}{4 p_i^2} + \frac{q^2}{4 p_j^2},
\end{gather}
with $\theta$ being the angle between $\mathbf p_i+\mathbf p_{i'}$ and $\mathbf p_j+\mathbf p_{j'}$.

Longitudinal and transverse polarization operators have the following form
\begin{gather}
    \Pi_l^{aa} = \sum_{i} (q_{l,i}^{a} )^{2} \chi_{l} + 4 (q_{C}^a )^{2} \chi_l,~~a=\{T_{8},Q\},    \\ \notag
    \Pi_t^{aa} = \sum_{i} (q_{t,i}^{a} )^{2} \chi_t + 4 (q_{C}^{a} )^2 \chi_t + \\ \notag 4 (q_{C}^a )^2 \chi_{sc},~~a=\{X, \tilde{Q}\},
\end{gather}
where the screening functions are used in the static limit
\begin{equation}
    \chi_l = 1, \qquad \chi_t = i \frac{\pi}{4} \frac{\omega}{q}, \qquad \chi_{\mathrm{sc}} = \frac{1}{3}
\end{equation}
The Debye masses are 
\begin{gather}
    (q^{a}_{l,i})^{2} = (Q^{a}_{i})^{2}\frac{p_i^2}{v_i\pi^2},~~~
    (q^{a}_{t,i})^{2} = (Q^{a}_{i})^{2}\frac{p_i^2}{\pi^2},\\ \notag
    (q^{a}_{C})^{2} = (Q^{a}_{C})^{2}\frac{\mu_{C}^{2}}{\pi^{2}}.
\end{gather}

\section{Thermal conductivity}

Thermal conductivity $\kappa_i$ in degenerate matter can be defined through characteristic relaxation time $\tau_i$ as 
\begin{gather}
\kappa_i = \tau_i \frac{\pi^2 T n_i }{3\mu_i}. 
\end{gather}

From the Boltzmann equation one can obtain the following expression for thermal conductivity
\begin{gather}
\kappa_i = 3 \tau_i \frac{(2\pi)^4}{T^2}\sum_{j}\nu_i \nu_j \sum_{1234} |M_{ij}|^2 f^0_1 f^0_2 (1-f^0_3)(1-f^0_4) \\ \notag 
\times \delta(\epsilon_\text{in}-\epsilon_\text{out})\delta(\mathbf p_\text{in}-\mathbf p_\text{out})[\tau_i(\phi_1-\phi_3)\phi_1+\tau_j(\phi_2-\phi_4)\phi_1]
\end{gather}
where $\sum_n \equiv \int d^3p/(2\pi)^3$, $f_n^0$ is Fermi-Dirac distribution function and $\phi_n \equiv (\epsilon_n-\mu_n)\mathbf v_n$, $\nu_i=2$ is a spin degeneracy of spin $1/2$ fermions.

Taking the limit $\omega,T\ll\mu_q$ we have
\begin{gather}
\kappa_i = \tau_i \frac{3}{16\pi^6}\sum_{j}\nu_i \nu_j \mu_i^2\mu_j^2 \int_0^\infty d\omega \left( \frac{\omega/(2T)}{\sinh(\omega/(2T))} \right)^2 \\ \notag 
\times \int_0^{q_M} dq \int_0^{2\pi} d\theta |M_{ij}|^2 [\tau_i(\phi_1-\phi_3)\phi_1+\tau_j(\phi_2-\phi_4)\phi_1],
\end{gather}
where $q_M = \text{min}[2p_1,2p_2].$

Since we consider degenerate matter in the limit $T\ll\mu_q$ we can perform the following substitution in the resulting formula:  $p_i \rightarrow\mu_i$ and $v_i\rightarrow1$ for light particles (i.e. electrons, $u$ and $d$ quarks);  $p_i \rightarrow\sqrt{\mu_i^2-m_i^2}$ and $v_i\rightarrow \sqrt{1-m_i^2/\mu_i^2}$ for massive particles (i.e. $s$ quarks). 

Finally, in the static limit we write
\begin{gather}
\tau_i(\phi_1-\phi_3)\phi_1+\tau_j(\phi_2-\phi_4)\phi_1 = \\ \notag
\frac{\omega^2}{2} \left[  \tau_i +\tau_j
\left(
\frac{q^{2}}{4 p_i p_j}
- \cos\theta
\sqrt{\left(1-\frac{q^{2}}{4p_i^2}\right)
\left(1-\frac{q^{2}}{4p_j^2}\right)}
\right) 
   \right].
\end{gather}

Gathering all equations, we get the closed system for the relaxation timescales $\tau_i$
\begin{gather}
1 = \frac{3}{4\pi^7 T} \frac{\mu_i^3}{n_i} \sum_{j} \mu_{j}^{2}
\left(\tau_{i}^{\kappa} s_{ij} + \tau_{j}^{\kappa} \tilde{s}_{ij} \right),
\end{gather}
where
\begin{gather}
s_{ij}^{\kappa}
= \int_{0}^{\infty} d\omega
\left(\frac{\omega/2T}{\sinh(\omega/2T)}\right)^{2}
\int_{0}^{q_{M}} dq
\int_{0}^{2\pi} \frac{d\theta}{2\pi}\,
|M_{ij}|^{2}\,\omega^{2},
\end{gather}

\begin{gather}
\tilde{s}_{ij}^{\kappa}
= \int_{0}^{\infty} d\omega
\left(\frac{\omega/2T}{\sinh(\omega/2T)}\right)^{2}
\int_{0}^{q_{M}} dq
\int_{0}^{2\pi} \frac{d\theta}{2\pi}\,
|M_{ij}|^{2}\\ \notag
\times \omega^{2}
\left(
\frac{q^{2}}{4 p_i p_j}
- \cos\theta
\sqrt{\left(1-\frac{q^{2}}{4p_i^2}\right)
\left(1-\frac{q^2}{4p_j^2}\right)}
\right).
\end{gather}

{\bf Acknowledgements.} We thank the anonymous referee for his important remarks, which substantially improved the presentation in the manuscript. This work is supported within the joint BRFFR-ICRANet-2025 funding programme under the grant No. F25ICR-001 and the National Natural Science Foundation of China under Grant No. 12275234.

\bibliography{total}{}
\bibliographystyle{h-physrev}


\end{document}